\documentclass[aps,epsfig,showpacs,superscriptaddress,preprint,12pt]{revtex4-1}
\usepackage{graphicx}
\usepackage{graphics}
\usepackage{color}
\usepackage{float}
\usepackage{epsfig}
\usepackage{textcomp}
\begin{document}
\title{The crowding effect on the melting of short DNA: Comparision with experiments}
\author{Neha Mathur}
\affiliation{Department of Physics, BITS Pilani, Pilani campus, 333031, Rajasthan, India}
\author{Amar Singh}
\affiliation{Center for Computational Biology, The University of Kansas - Lawrence, KS 66047, USA}
\author{Navin Singh}
\affiliation{Department of Physics, BITS Pilani, Pilani campus, 333031, Rajasthan, India}
\email{navin@pilani.bits-pilani.ac.in}
\date{\today}

\begin{abstract}
We study the effect of crowders on the melting profile of homogeneous and heterogeneous DNA molecules. We find out the melting profile of short DNA molecules and compare our findings with the experiments. We consider some random distribution of crowders along the chain, and by finding out the best match with the experiments, we attempt to identify the location of crowders in the experimental findings of Ghosh \cite{Ghosh_PNAS_2020}. We also study the melting of homogeneous DNA molecules of different lengths (25, 50, 75) in the presence of only one crowder in the chain. By varying the location of the crowder from one end to the other, we find that the melting temperature is susceptible to the location of the crowder at the ends. At the same time, there is minimal effect on the melting temperature due to the location of the crowder. {\it In vivo}, the strength of a crowders may vary along the chain. We study the melting of long heterogeneous chain in presence of five crowders of different strength. We find that there is a significant variation in the melting process of DNA in presence of crowders of variable strength. 
\end{abstract}

\maketitle

\section{Introduction}

The presence of various biomolecules, \textit{for eg} proteins, nucleic acids, saccharides, lipids, and metabolites, makes the cell a crowded environment \cite{Rivas_Biophy_2018,Kim_SoftMatter_2015}. It is known that these molecules occupy about 20–40\% of the space in the cell \cite{Akabayov_Nat_2013}. In the recent years, researchers are investigating the role of the crowders on various biological activities of the biomolecules. All these works reveal many interesting features of activities of the biomolecules under the influence of molecular crowders \cite{Kumar_JSM_2017, Taylor_JCP_2020,Jung_JCP_2021}. The effect of these crowders has an essential effect on molecular transport, reaction rates, and chemical equilibrium. These crowders also affect the condensation of DNA, the cell surface, and the interior of the cell \cite{ZimmerFEBSL_1996,Hormen_Biopol_2012,Khimji_Chemc_2013}. The molecular crowders play a crucial role in phase separation in the cytoplasm and condensation of DNA into the nucleoid of bacterial cells \cite{Ghosh_NAR_2019}. Recently, Takanashi has shown how the molecular crowders modify the DNA and RNA polymerase reactions \cite{Takahashi_Mol_2020,Takahashi_RSC_2020}. In another interesting study, Ghosh {\it et al.} investigated the stability of short DNA duplexes in different solutes using the nearest neighbour (NN) model \cite{Ghosh_PNAS_2020}.  The excluded volume due to the crowders significantly affects the solution's osmotic pressure. Despite many interesting results, the molecular crowding effects on the properties of biomolecules are still unclear. Out of these activities, the melting of DNA is an interesting area of research \cite{Amar_PCCP_2017,Zoli_PCCP_2019}. The transformation of double-stranded DNA (dsDNA) to two single-stranded DNA (ssDNA), due to thermal fluctuation or due to a force or due to the change in the pH of the solution, is known as melting of DNA \cite{Wartell_PR_1985,Kumar_PR_2010}. 

In the present work, we study the melting of DNA molecule in the presence of molecular crowder(s) in thermal ensemble. We use the Peyrard-Bishop-Dauxois (PBD) model for the current investigation \cite{Peyrard_PRL_1989,Dauxois_PRE_1993}. The objectives of this manuscript are as follows. First, to identify the location of crowders in the vicinity of DNA molecules for which the experimental data are available \cite{Ghosh_PNAS_2020}. Second to understand the melting of DNA, due to different locations of a single crowder. In contrast to the real scenario, most of the theoretical studies consider constant strengths of molecular crowders that are present in the cell. That is the objective of the current work, to investigate the thermodynamics of DNA molecule in presence of crowders of variable strengths. The manuscript is divided into four sections: in Sec. \ref{model}, we explain the modifications in the PBD model to take care of the molecular crowder. The melting of short DNA and comparison with the experiments is discussed in Sec. \ref{short}. We discuss the effect of location of single crowder on the melting of DNA in Sec. \ref{inter}. In sec. \ref{long}, we discuss the melting of long DNA chain in presence of crowders of variable strengths. We summarize our findings and importance of the obtained results in Sec. \ref{conc}.

\section{Model}
\label{model}

To study the effect of molecular crowders on the melting of DNA molecules in the thermal ensemble, we use the well known Peyrard-Bishop-Dauxois model (PBD). The model is quasi-one-dimensional and expresses the dynamics of the molecule through the stretching of the hydrogen bonds \cite{Peyrard_PRL_1989,Dauxois_PRE_1993}. The model underestimates the entropy associated with the different conformations of the molecule. The linear form of the model ignores the effect of the molecule's helicoidal nature and the solvent effect of the solution. Despite these shortcomings, the model still has enough details to describe the DNA molecule denaturation/unzipping process. The interactions in the DNA, containing $N$ base pairs, are represented as,
\begin{equation}
\label{eqn1}
H = \sum_{n=1}^N\left[\frac{p_n^2}{2m}+ V_m(y_n) \right] + \sum_{n=1}^{N-1}\left[V_s(y_n,y_{n-1})\right],
\end{equation}
here $y_n$ represents the separation between two bases in a pair. The separation $y_n = 0.0$ \AA{} refers to the equilibrium position of two bases in a pair. First term of the model is the momentum term which is $p_n = m\dot{y}_n$. We have taken same reduced mass, $m = 300 $ amu for both the $AT$ and $GC$ base pairs \cite{Peyrard_PRL_1989}. The interaction between the nearest base pairs along the chain, the stacking interaction, is represented by, 
\begin{equation}
\label{eqn2}
V_s(y_n,y_{n-1}) = \frac{\kappa}{2}(y_n - y_{n-1})^2[1 + \rho e^{-b(y_n + y_{n-1})}].
\end{equation}
The single-strand elasticity is represented by $\kappa$, representing the anharmonicity in the strand elasticity by $\rho$. The parameter, $b$, describes the range of anharmonicity. The earlier works demonstrate that values of $k$ and $\rho$ defines the sharpness in the transition from double-strand to single strand \cite{Cocco_PRL_1999, Navin_EPJE_2005,Zoli_PCCP_2020}. The Morse potential represents the hydrogen bond between the two bases in the $i^{\rm th}$ pair.
\begin{equation}
\label{eqn2a}
V_M (y_n) = D_n(e^{-a_ny_n} - 1)^2. 
\end{equation}
where $D_n$ represents the potential depth, and $a_n$ represents the inverse of the width of the potential well. These two parameters have a crucial role in DNA denaturation. From previous results, we know that the bond strengths of these two pairs are in an approximate ratio of $1.25$ - $1.5$ as the GC pairs have three while AT pairs have two hydrogen bonds. The complete set of parameters is, $D_{\rm AT}$ = 0.0395 eV, $D_{\rm GC}$ = 0.059 eV, $a_{\rm AT}$ = 4.2 ${\rm \AA^{-1}}$, $a_{\rm GC}$ = 6.3 ${\rm \AA^{-1}}$, $\rho$ = 2.0, $\kappa$ = 0.03 ${\rm eV/\AA^{2}}$, $b$ = 0.35 ${\rm \AA^{-1}}$. The model parameters are tuned in such a way we get a good match with the experimental results. We can study the thermodynamics of the transition by evaluating the partition function. For a sequence of $N$ base pairs, the canonical partition function can be written as:
\begin{equation}
\label{eqn3}
Z = \int \prod_{i=1}^{N}\left\{dy_idp_i\exp(-\beta H)\right\} = Z_pZ_c,
\end{equation}
where $Z_p$ corresponds to the momentum part of the partition function and is equal to $(2\pi mk_BT)^{N/2}$. The configurational part of the partition function, $Z_c$, is defined as, 
\begin{equation}
\label{eqn4}
Z_c = \int_{-\infty}^{\infty}e^{-\frac{\beta V(y_1)}{2}}dy_1 \prod_{i = n}^{N-1} dy_n e^{-\frac{\beta}{2}[V(y_n) 
+ V(y_{n+1})+ 2 W(y_n, y_{n+1})]} e^{-\frac{\beta V(y_N)}{2}} dy_N
\end{equation}

We adopt the following method to calculate the partition function for the chains with a random sequence of $AT$ and $GC$ pairs and open boundaries. The partition function in the PBD model is divergent, a proper cut-offs are required to overcome this issue. From our previous studies, we conclude that an upper cut-off of 200 ${\rm \AA}$ is sufficient to overcome the divergence issue of the partition function. The lower limit of integration is set as -5.0 \AA{} \cite{Zhang_PRE_1997,Erp_EPJE_2006,Amar_PRE_2015,Navin_EPJE_2005}. Once we find the proper cut-offs, the task is to discretize the integral in Eq.\ref{eqn4}. We use the Gaussian quadrature to integrate the equation of partition function numerically. We discretize the configurational space into 900 points. Once we can evaluate the partition function, we can determine the thermodynamic quantities of interest by evaluating the Helmholtz free energy of the system. We define the Helmholtz free energy per base pair as,
\begin{equation}
\label{eqn5}
f(T) = -\frac{1}{2}k_B T\ln\left(2\pi m k_B T\right) - \frac{k_B T}{N}\ln Z_c
\end{equation}
In the thermal ensemble, the specific heat, $C_v$, is evaluated by taking the second derivative of the free energy, as $C_v = -T(\partial^2 f/\partial T^2)$. We calculate the chain's melting temperature ($T_m$) from the peak in the specific heat curve. In the experiments, researchers monitor the fraction of open pairs, $\phi$, as a function of temperature using various spectroscopic techniques. To calculate the $\phi$, we adopt the method as discussed by Campa \cite{Campa_PRE_1998}. We calculate the value of $\theta$, the average fraction of open pairs as a function of temperature. The average fraction of bonded base pairs, $\theta(= 1 - \phi)$ is defined as \cite{Campa_PRE_1998,Navin_PRE_2001}.
\begin{equation}
\label{eq_theta}
\theta = \theta_{\rm ext}\theta_{\rm int} 
\end{equation}
$\theta_{\rm ext}$ is the average fraction of strands forming duplexes. The $\theta_{\rm int}$ is the average fraction of unbroken bonds in the duplexes. To compute $\theta_{\rm int}$, one has to separate the configurations describing a double strand on the one hand and dissociated single strand on the other. The $n^{th}$ bond is considered to be broken if the value of $y_n$ is larger than a cutoff $y_0$. One can therefore define $\theta_{\rm int}$ as:
\begin{equation}
\label{eq_theta_int}
\theta_{\rm int} = \frac{1}{N} \sum_{n=1}^{N}\langle \vartheta(y_0 - y_n)\rangle
\end{equation}
where $\vartheta(y)$ is Heaviside step function and the canonical average $\langle.\rangle$ is defined considering only the double strand configurations. To compute $\theta_{\rm ext}$ we use the method discussed in \cite{Campa_PRE_1998} :
\begin{equation}
\label{eq_theta_ext}
\theta_{\rm ext} = 1 + \delta - \sqrt{\delta^2 + 2\delta}
\end{equation}
where
\begin{equation}
\delta = \frac{Z(s_1)Z(s_2)}{2N_0 Z(ds)} 
\end{equation}
here $Z(s_1)$, $Z(s_2)$ and $Z(ds)$ are the configurational isothermal isobaric partition functions of systems consisting of molecular species single strand $s_1$, single strand $s_2$ and the double-strand configuration dsDNA respectively. The internal and external parts of the partition function are,
\begin{equation}
\frac{Z(s_1)Z(s_2)}{2N_0 Z(ds)} =  \frac{Z_{int}(s_1)Z_{int}(s_2)}{a_{av}Z_{int}(ds)} 
\frac{a_{av}Z_{ext}(s_1)Z_{ext}(s_2)}{2N_0 Z_{ext}(ds)} 
\end{equation}
where $a_{av.} = \sqrt{a_{AT}a_{GC}}$.
In analogy to what has been proposed for the Ising model on the basis of partition function of rigid molecules, one makes the following choice \cite{Campa_PRE_1998}
\begin{equation}
\frac{a_{av.}Z_{ext}(s_1)Z_{ext}(s_2)}{2N_0 Z_{ext}(ds)} = \frac{n^*}{n_0}N^{-p\theta_{\rm int} + q} 
\end{equation}
and then,
\begin{equation}
\delta = \frac{1}{a_{av}}\frac{Z_{int}(s_1)Z_{int}(s_2)}{Z_{int}(ds)}   \frac{n^*}{n_0}N^{-p\theta_{\rm int} + q} 
\end{equation}
where $n^*$ is a chosen reference concentration as 1$\mu M$ while $n_0$ is the single strand concentration which we have chosen as 3.1 $\mu M$. $p$ and $q$ are the parameters which can be calculated using experimental results \cite{Campa_PRE_1998,Navin_PRE_2001}. We have taken the parameters $p$ and $q$, as $p=29.49$ and $q=27.69$. Using this equation we calculate the $\theta_{\rm ext}$ and hence the fraction of open pairs, $\phi$ as a function of temperature.
\begin{figure}[ht]
\begin{center}
\includegraphics[height=2.5in,width=3.2in]{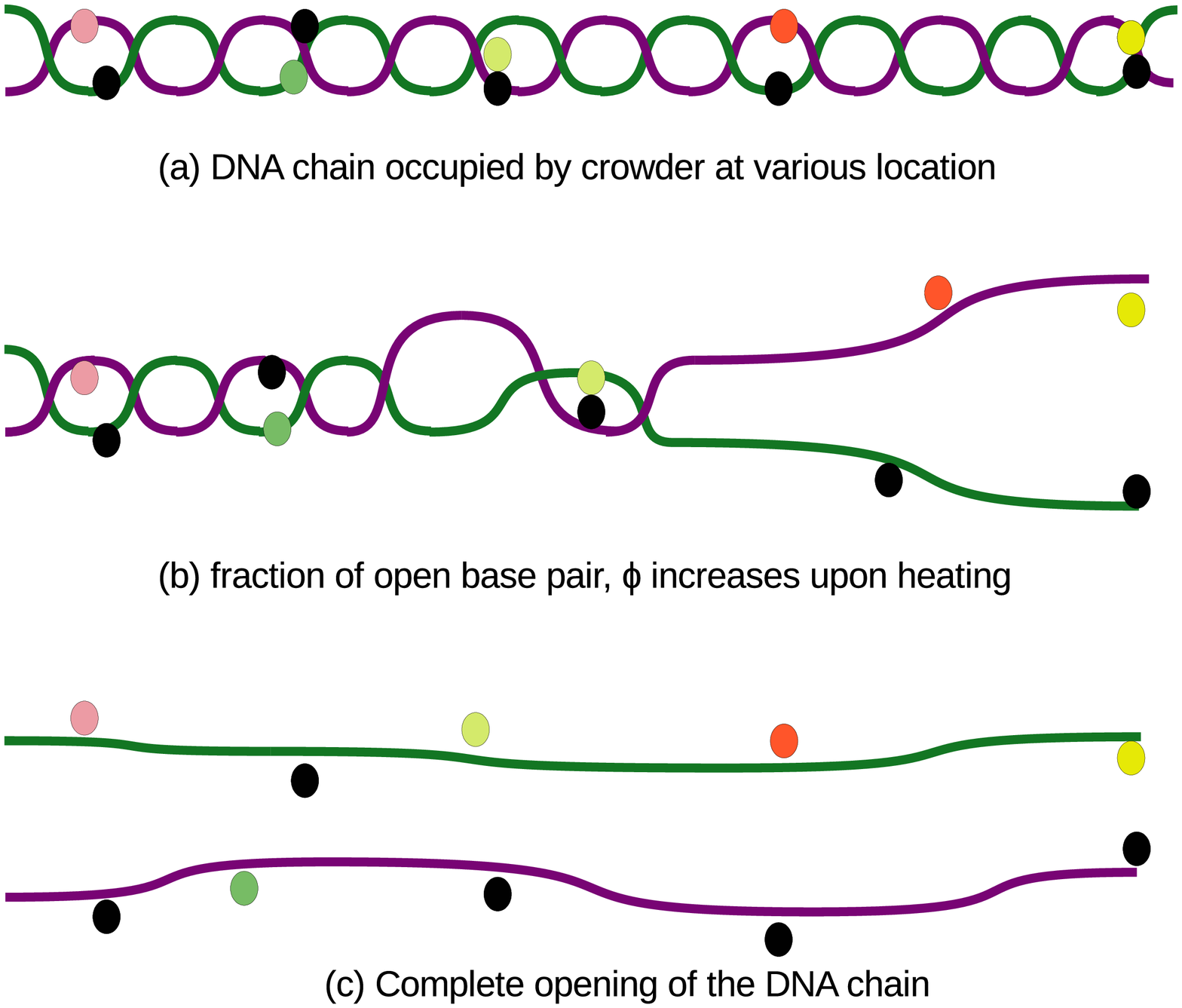}
\includegraphics[height=2.5in,width=3.0in]{fig01b.eps}
\caption{\label{fig01} (a) The schematic represention of the crowders (shown with coloured circle) present in the surrounding of DNA molecule. (b) The realization of crowder in the model. We modify the depth of the potential for the base pair surrounded by a crowder as $D = \alpha*D_0$, where $\alpha$ is a scaling factor.} 
\end{center}
\end{figure}

\section{Melting of short DNA}
\label{short}

We consider the DNA molecule considered by Ghosh {\it et al} \cite{Ghosh_PNAS_2020}. They considered 40\% PEG 200 in 100mM Nacl in order to form the crowder environment in the surrounding of DNA. In the present problem, we have two kinds of base pairs in the chain: one that is not having any crowder in the surrounding and other having crowder(s) in the surrounding. Our approach is theoretical, we need certain modifications in the model parameters or terms. We propose the following argument. A base pair that is surrounded by a crowder needs a very high energy to break the hydrogen bond. In the PBD model, the base pairs can move only along the $y$ direction, thus the space available to the base pair surrounded by a crowder will be restricted. The crowders are the biomolecules which can move due to the thermal fluctuation in the DNA molecule. To the zeroth approximation, we can argue that the base pair, that is surrounded by a crowder, requires very high amount of energy to overcome the potential barrier or to break the hydrogen bonds. We, therefore, modify the depth of potential as $D = \alpha*D_0$, where $\alpha$ is the scaling factor. We take $\alpha=1.5$ for the site where the crowder is present \cite{Amar_PCCP_2017}. 
\begin{figure*}[ht]
\begin{center}
\includegraphics[height=5.0in,width=6.5in]{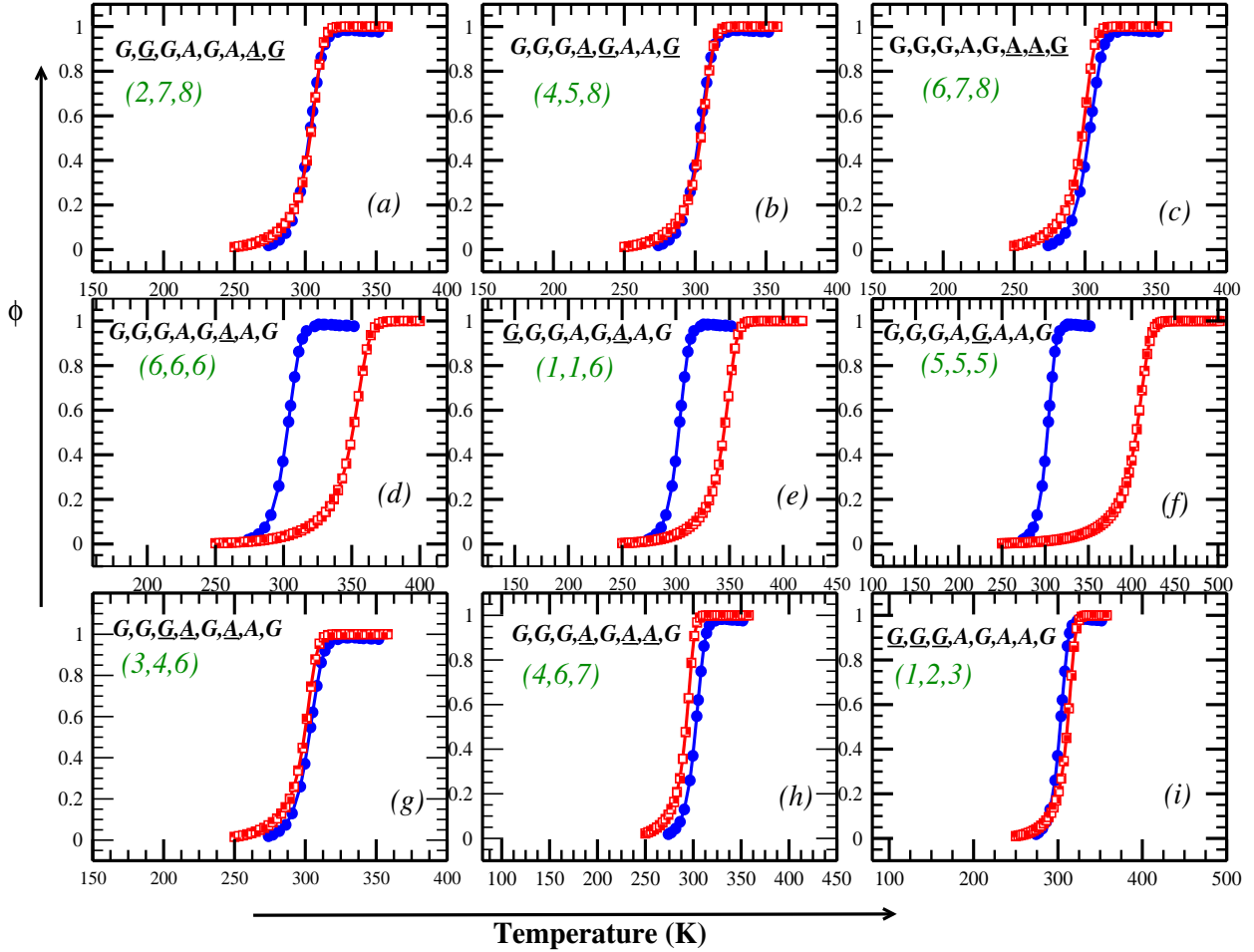}
\caption{\label{fig02} The change in the fraction of open base pairs with temperature for chain-A ( $G,G,G,A,G,A,A,G$). In this figure red line corresponds to results from the PBD model while blue colour corresponds to the experiment findings. The underlined letters are the sites where the crowder is present.} 
\end{center}
\end{figure*}
First we consider the chains for which the experimental data are available \cite{Ghosh_PNAS_2020}.
The sequences are: (chain-A) $5'-G,G,G,A,G,A,A,G-3'$ and (chain-B) $5'-G,G,A,A,G,A,G,G-3'$. In the experiments, they have considered 40\% of the volume occupied by crowders which means for the eight bases pairs chain there will be three crowders located at the random locations in the vicinity of the DNA. Our purpose of the current investigation is to find out the best fit with the experimental results which can help us in identifying the location of crowders in the experimental findings. Hence we consider different locations of the crowders and investigate the effect of crowders and its location on the melting profile of DNA molecule. The results are shown in figures \ref{fig02}($a-i$).
\begin{figure*}[h]
\begin{center}
\includegraphics[height=4.5in,width=5.5in]{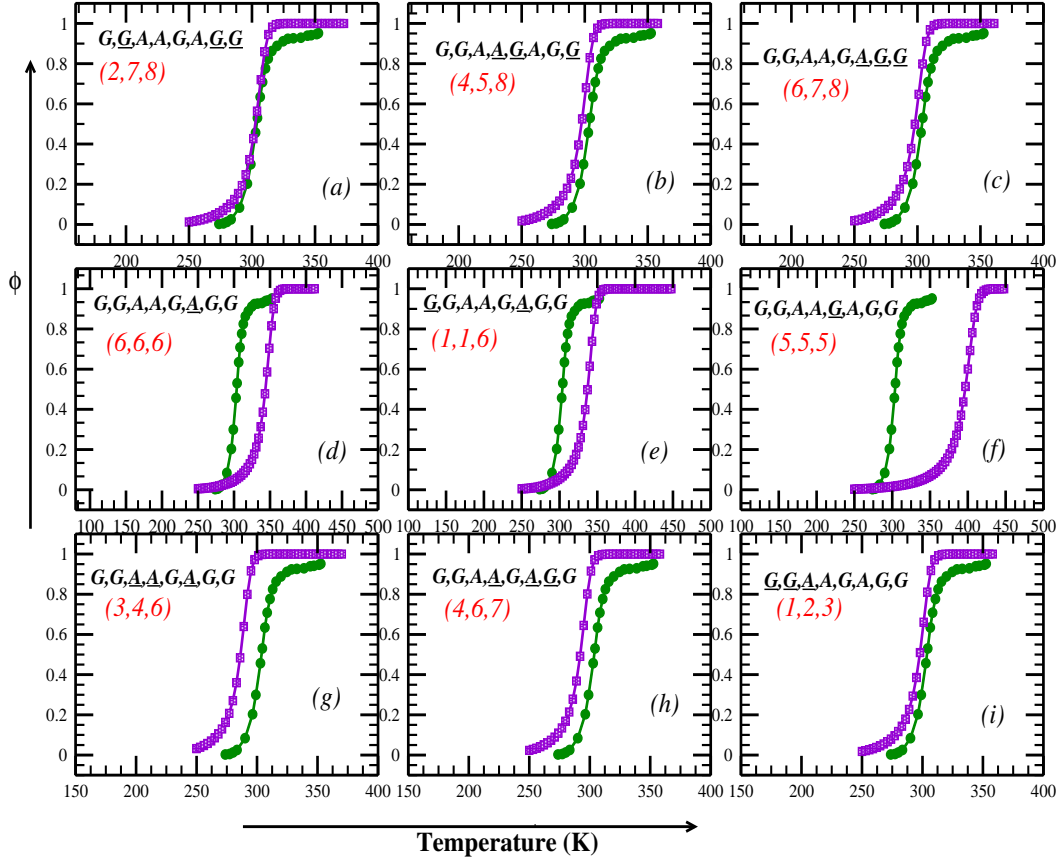}
\caption{\label{fig03} The change in the fraction of open base pairs with temperature for chain-B ( $G,G,A,A,G,A,G,G$). In this figure purple line corresponds to results from the PBD model while green colour corresponds to the experiment findings. The underlined letters are the sites where the crowder is present.} 
\end{center}
\end{figure*}
In figure \ref{fig02}a, the crowders are present at $2^{\rm nd}, 7^{\rm th}, 8^{\rm th}$ sites. The distribution of crowders in this case, restricts the chain to open from the ends and the chain is forced to open from mid region. Similarly we place the crowders on other locations and find out the match with the experiments. Among various possible distributions, we are getting the close match for the distribution ($2,7,8$) or ($4,5,8$). The results are indicating that in the experiments, the crowders might be present in the such a way that two crowders are located in one half of the chain while the third crowder is present at the other end of the chain. The results are very sensitive to the loctation of the crowders as apparent from the rest of the figures. In the experiments it is difficult to fix the crowder's location, hence we also explore the possibility of one site having more than one crowder (see fig. \ref{fig02}d-f). Our results are indicating that in the experiment the crowders are distributed through out the chain. To check whether our results are sensitive to the sequence or not, we consider another sequence (chain-B) for which the experimental results are available \cite{Ghosh_PNAS_2020}. Again we have considered large number of distributions of crowders in the chain. Here we are showing the results for only nine cases. The results are plotted in fig. \ref{fig03}. From the figures, we can predict that when the crowders are at ($2,7,8$) there is a close match with the experiment. The results again indicate that the location of crowders are very important and the melting profile is very sensitive not only to the sequence but also to the location of crowders. For the chain-B we are not getting good match for the distributions ($4,5,8$) for which we got a good match for chain-A. Point to note that both the chains are having $5GC$ and $3AT$ pairs and the crowded sites are having same kind of pair ($A,G,G$). However, there is a difference in the sequence of the base pairs in the other half of the chain where there is no crowder. The chain-A is having $5'-G,G,G,A$ pairs while the chain-B is having $5'-G,G,A,A$ pairs where there is no crowder. This means that chain-B is little more entropic than chain-A. We can say that this is due to limitation of the model which is unable to capture the minor details of the experiment and hence there is slight mismatch with the experimental results. 

\section{Crowder's location and melting of DNA molecule}
\label{inter}

In this section, we discuss the effect of molecular crowders on the DNA molecules of different lengths. We consider a homogeneous chain of 25, 50 and 75 base pairs and find out the melting profile of these DNA molecules in presence of single crowder in surrounding of the DNA. Our aim of this study is to visualize the effect of crowder and its location on the melting profile of DNA molecule.
\begin{figure}[t]
\includegraphics[height=2.75in,width=3.5in]{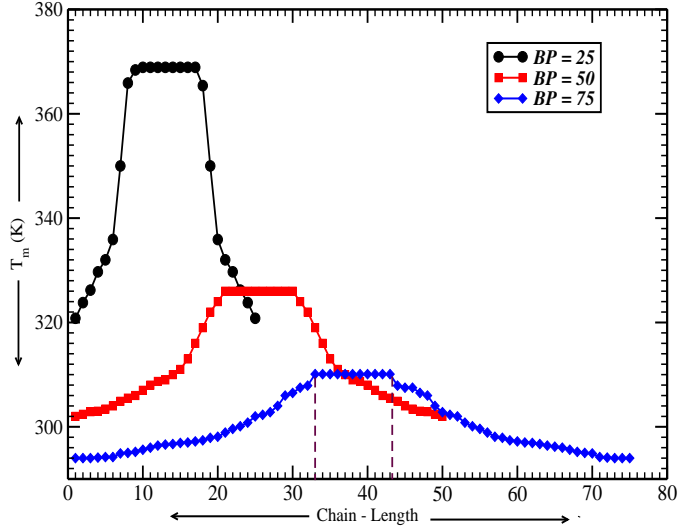}
\caption{\label{fig04} \small The melting temperature of DNA molecule of $25,\; 50$ and $75$ base pairs in the presence of one crowder. The fig (a) shows a comparison of melting temperature of all chains. The figures (b), (c) and (d) are for the DNA molecule of 25, 50 and 75 bases pairs respectively. The melting temperatures of $25,\; 50$ and $75$ are 360.0 $K$, 320.3 $K$ and 305.0 $K$ respectively.}
\end{figure}
We place the one crowder at a time in the surrounding of a base pair in the DNA molecule. Using the method discussed above, we calculate the melting temperature of the molecule. To study the effect of crowder's location on the melting of DNA chain, here we have taken little larger scaling factor. We take $\alpha = 10$. In a similar fashion we calculate the melting temperature of the DNA molecule by moving the location of crowder from one end to the other end. We calculate the melting temperatures for all the three chains in the same manner. The results are shown in the fig. \ref{fig04}. We are getting two important results from this study. First, there is an overall decrease in the $T_m$ with the increase in the chain length. Second, for all the three chains while the melting temperature is very sensitive to the location of crowder when it is near the ends, there is a region, in the middle of the chain, where the location of crowder is not important. There is a region of approximately ten base pairs in the middle of the chain where the location of crowder is not important, irrespective of the chain length of the DNA molecule. We know that the DNA molecule opens either from the middle, in form of a bubble, or from the ends due to the end entropy. When a crowder is placed near the ends, there is a reduction in the end entropy which is sensitive to the location of the crowder. The same argument is not true when the crowder is present somewhere in the middle of the chain. We can see that there is no change in the melting temperature of the molecule due to the change in the location of crowder for about ten base pairs in the middle section of the chain (see fig. \ref{fig04}).

\begin{figure*}[h]
\includegraphics[height=2.5in,width=3.in]{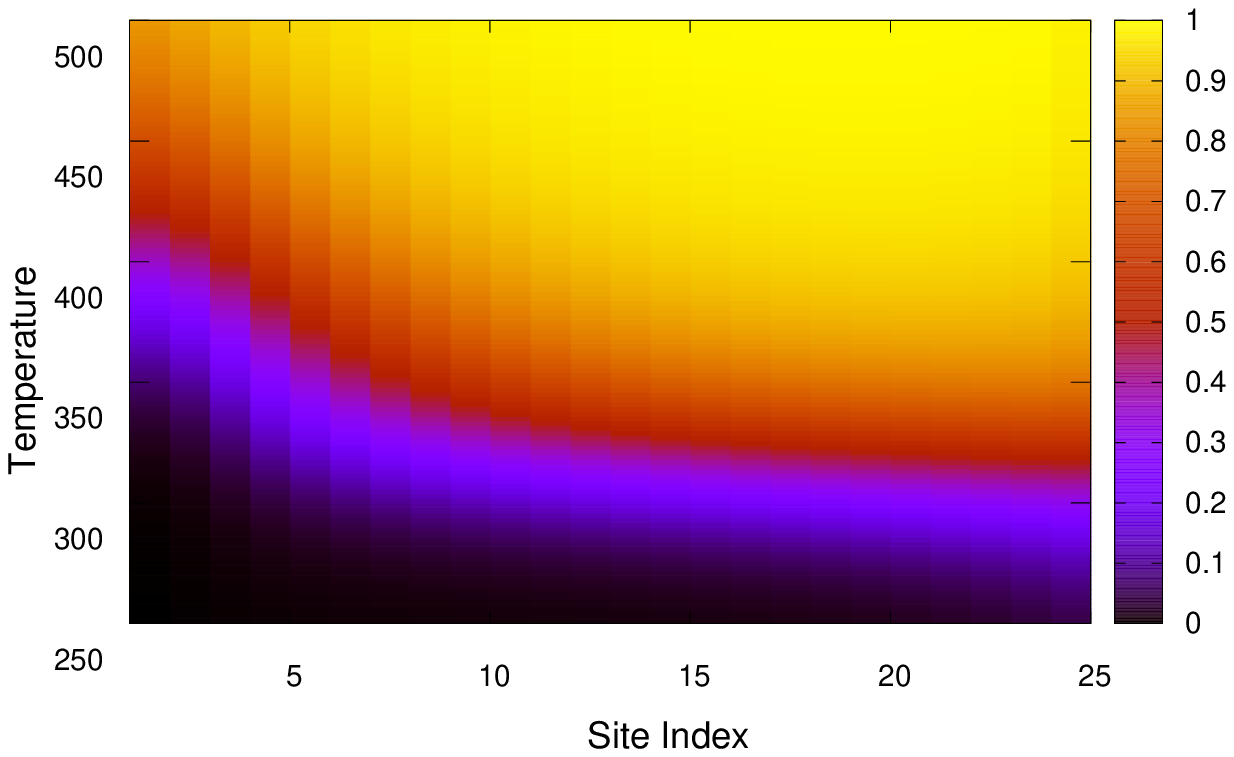}
\includegraphics[height=2.5in,width=3.in]{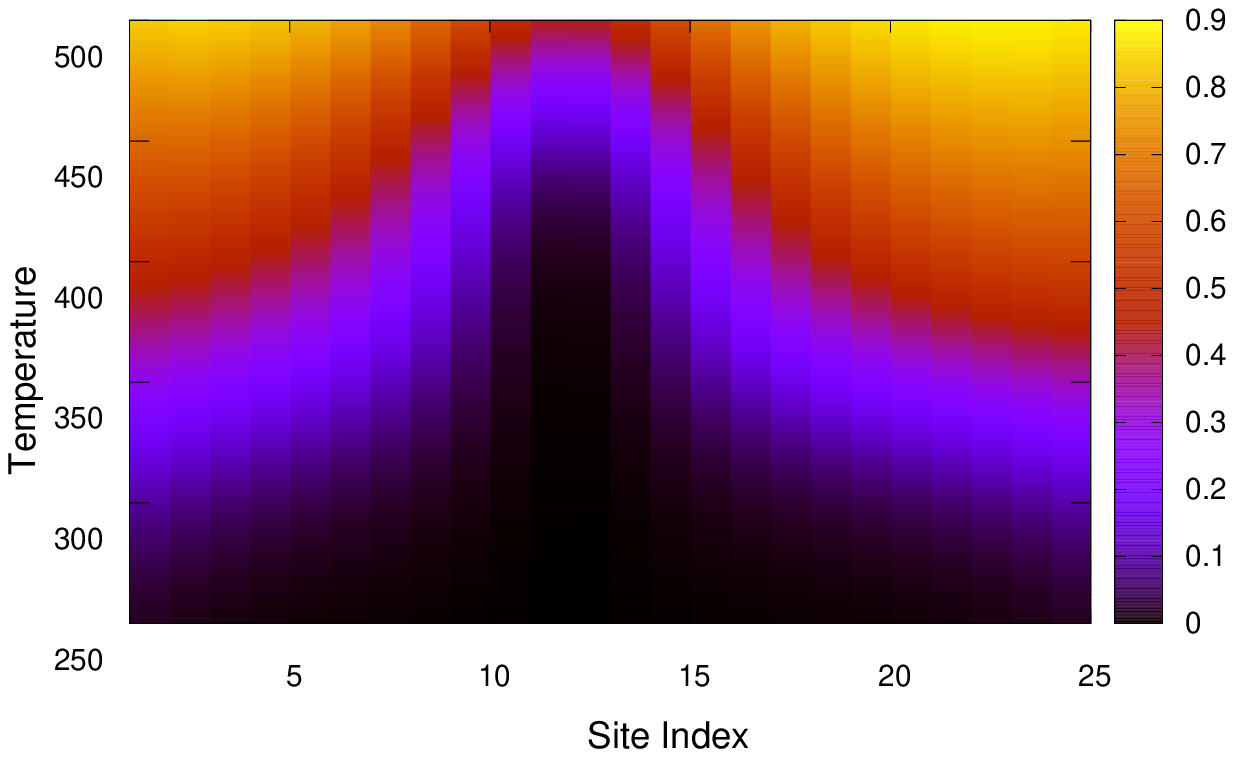}
\caption{\label{fig05}\small \\ The opening profile of 25 base pair DNA molecule in the presence of crowder at the (a) end and (b) in the middle of the chain. To avoid the overflow of the plots, we are showing the results for 25 base pairs only.}
\end{figure*}

To obtain an overview of the melting of the DNA of different lengths we calculate the probabilities of opening of each pair of a DNA molecule. The probability of opening of the $i^{\rm th}$ pair, in a sequence is defined as \cite{Navin_JCP_2011}:
\begin{equation}
\label{eqn7}
P_j = \frac{1}{Z_c}\int_{y_0}^{\infty} dy_j \exp\left[-\beta H(y_j, y_{j+1})\right] Z_j
\end{equation}
where 
$$ Z_j = \int_{-\infty}^{\infty} \prod_{i=1, i\neq j}^N dy_i \exp\left[-\beta H(y_i, y_{i+1})\right]
$$
while $ Z_c $ is the configurational part of the partition function defined as in eq. (\ref{eqn7}). For $y_0$, we have taken a value of 2 ${\rm \AA}$. The opening profile of 25 base pair chain is shown in the fig.\ref{fig05}.\\

\section{DNA in the presence of more than one crowder}
\label{long}

\begin{figure*}[hbt]
\includegraphics[height=2.75in,width=4.in]{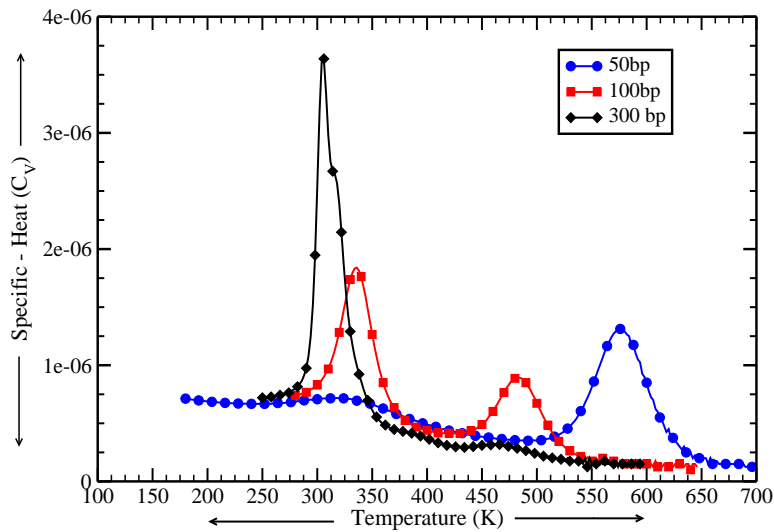}
\caption{\label{fig06} \small The change in the specific heat with temperature of 50, 100, and 300 base pairs DNA molecule in the presence of five crowders. There is a significance difference in the melting of the three chains in the presence of the crowders.}
\end{figure*}

We now consider the DNA molecules of a heterogeneous sequence of $AT$ or $GC$ pairs and study the molecule's stability in the presence of five crowders. In this part of the study, we aim to understand more about the effect of crowding on the stability of DNA molecules. In the real system, we have minimal information about the distribution of crowders and their strengths inside the cell. The distribution is heterogeneous and random. In order to be closer to the real system, we consider the random location of crowders of different intensities. Since we represent the crowder by the depth of Morse potential, we choose different scaling factor for each crowder. We choose, $\alpha = 5, 6, 7, 8, 9$. All other parameters are same as taken in the previous part of the problem. We consider the DNA molecule of three lengths: 50, 100 and 300 base pairs. The crowders location in the 50 base pair chain is 33, 36, 27, 15, and 43, while in the 100 base pairs chain, it is: 83, 86, 77, 15, and 93. Since the crowder's intensities are different for different sites, the crowded site is important. For example, for the 100 base pairs chain, the potential depth for $83^{\rm rd}$ site will be $5D_0$. For $86^{\rm th}$ site it will be $6D_0$, for $77^{\rm th}$ site it will be $7D_0$, for $15^{\rm th}$ site it will be $8D_0$ while for $93^{\rm rd}$ site it will be $9D_0$. We calculate the melting temperature of the system, the result of which is shown in fig. (\ref{fig06}). Interestingly the number of peaks and their locations are different in all three chains. While for the 50 base pairs chain, the first peak is of lower height than the second peak, for the 100 base pair molecule, the second peak is of lower height. For the 300 base pairs molecule, the crowders are present at 283, 286, 177, 115, and 293 sites. In this case, the second peak almost disappears. The results obtained here indicate that crowders in the cell affect the melting of the molecules of different lengths in a different fashion. In comparison of all the three chains, the 300 base pairs chain is highly entropic in nature.\\
\vskip 0.1in

\begin{figure*}[hbt]
\includegraphics[height=2.25in,width=2.in]{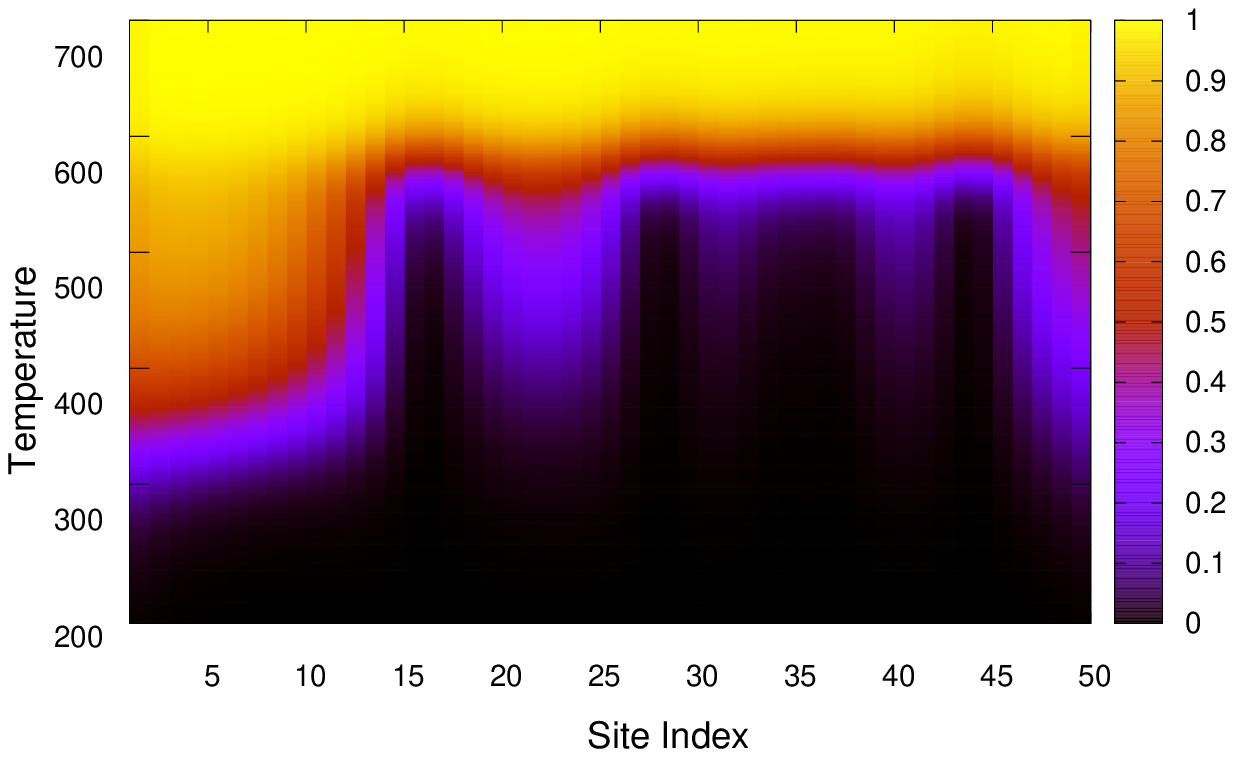}
\includegraphics[height=2.25in,width=2.in]{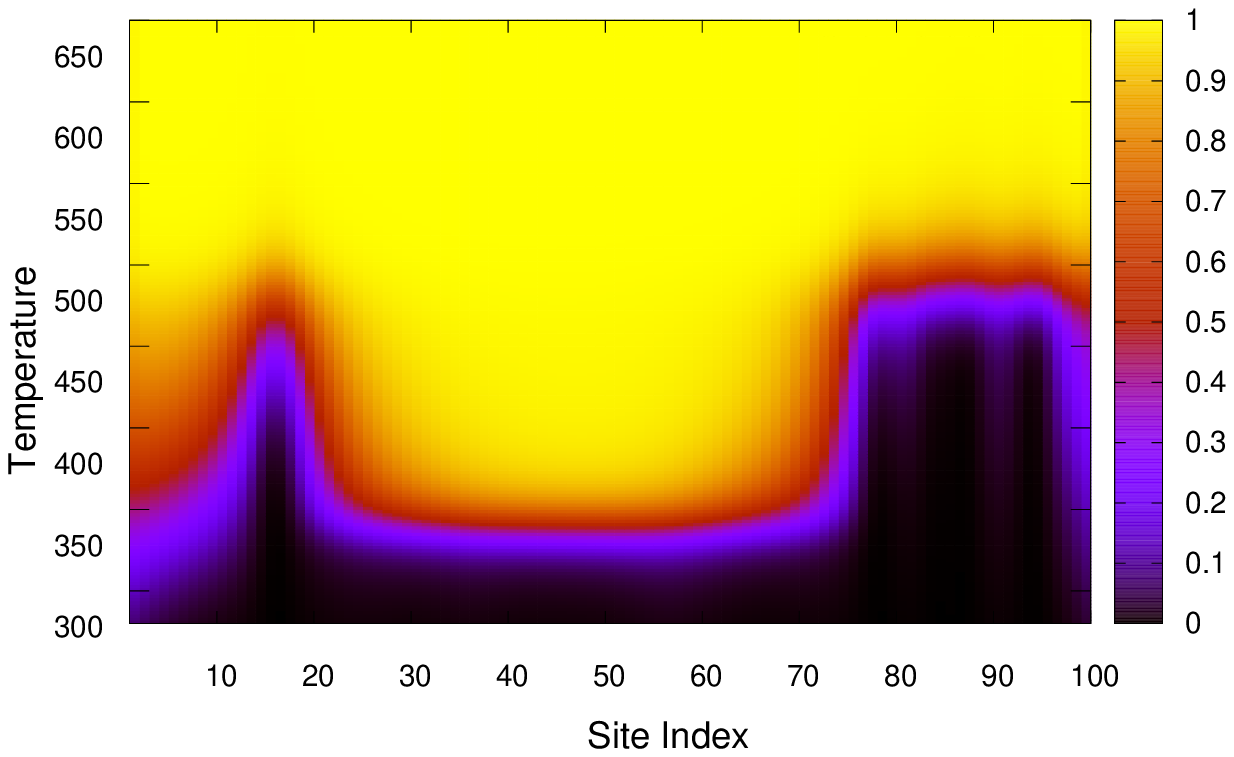}
\includegraphics[height=2.25in,width=2.in]{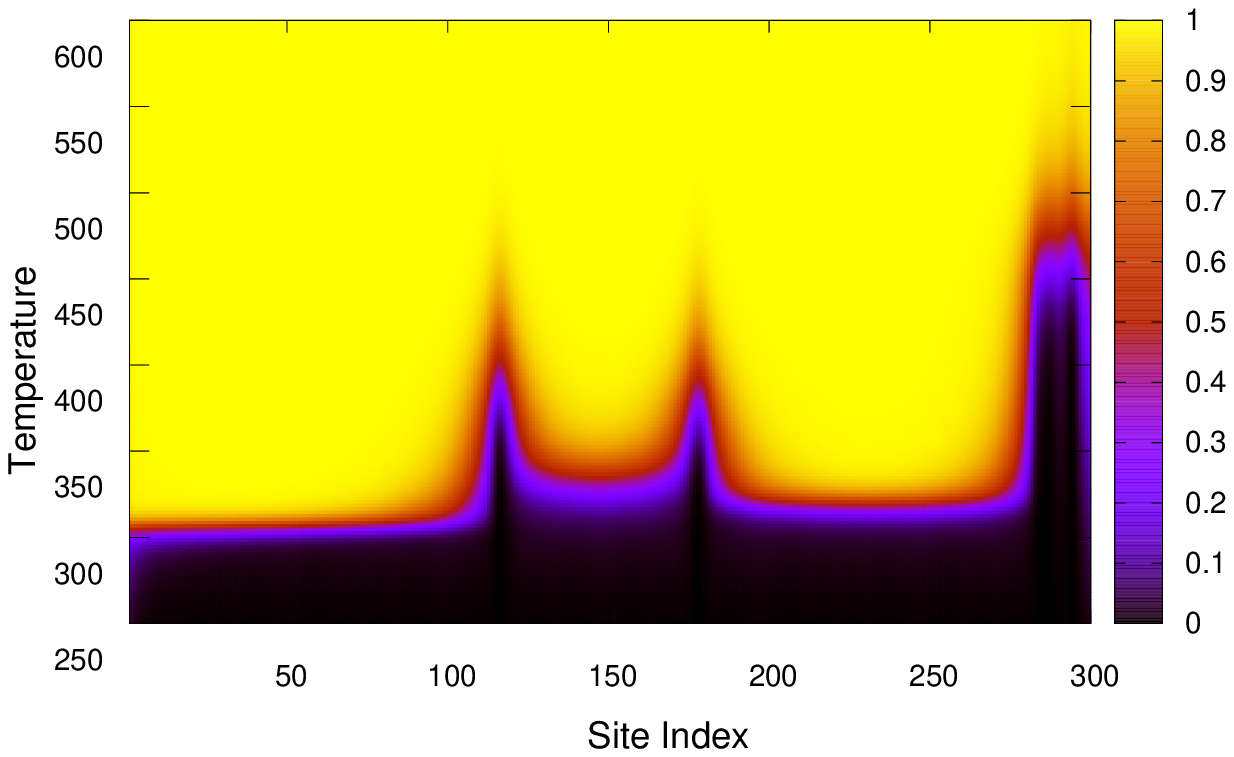}

\caption{\label{fig07} \\ \small The opening probabilities of the 50, 100, and 300 base pairs chain in the presence of five crowders. The location as well as the strength of the crowders are visible through the heights of the black region.}
\end{figure*}

To understand more about the opening of the molecule, we calculate the opening probabilities of all the three chains with the help of eq. (\ref{eqn7}). Fig. \ref{fig07} displays the obtained results. The three plots tell the story of the opening of the base pairs in all three chains. As discussed above, for the 50 base pairs chain, only a few base pairs (10-15) open initially (the reason for the first smaller peak). The major section of the chain opens nearly in a narrow temperature range. For the 100 base pairs chain, while a big bubble forms as the base pairs 20-70 (nearly 50\%) are in the open state, the ends are intact due to the presence of the crowder. The system needs higher energy to open the bonds at the ends. That is why we have the first peak higher than the second one in this case. In the 300 base pairs chain, there is a large bubble (200-275) and a larger opening at one end (1-95). In addition to these, there is a smaller bubble from 125-175 in the middle section of the chain. That is the reason for the high entropy of the 300 base pairs chain. Another important reason for the 300 base pairs chain to be more entropic is reducing the fraction of crowded sites. For 50 base pairs chain it is 10\%, 100 base pairs chain it is 5\%, while, for the 300 base pairs chain it is $\sim1.7\%$.

\section{Conclusions}
\label{conc}

We have studied the effect of crowders on the melting profile of homogeneous and heterogeneous DNA molecules. In the first part of the study, we have taken the DNA molecule of 8 base pairs, for which the experimental results are available. We have considered six different cases of random distribution of crowders along the chain. The purpose of this part of the study is to identify the location of crowders in the experimental case. {\it In vitro}, the melting temperature of the DNA molecule in the presence of the crowders is calculated. However, we do not have information about the location or distribution of the crowders. With suitable modifications in the PBD model for the sites where crowders are present, we have calculated the melting temperature for all six cases for both chains. We have found the best match with some of the experimental results. Through the obtained results or match, we can identify the probable location of the crowders in the chain.

In the second part of the study, we have studied the melting of homogeneous DNA molecules of different lengths (25, 50, 75). In this part of the study, we have taken only one crowder in the chain and have calculated the melting temperature of the system by changing the location of the crowder from one end to the other. We have found some interesting outcomes of this study. The melting temperature is susceptible to the location of the crowder at the ends. In contrast, there is a region of $\sim15$ base pairs in the middle, where the location of the crowder has minimal effect on the melting temperature of the system. The melting of the system is smooth as there is only one crowder in the chain, as depicted through fig. \ref{fig05}.

In the third part of the study, we have studied the melting of heterogeneous DNA molecules of different lengths (50, 100, 300) in the presence of 5 crowders of different strengths. Since {\it in vivo}, the crowder present in the surrounding of a base pair may have a different strength than the crowder present in the surrounding of any other base pair along the chain. The very interesting outcome of this part of the study is the presence of multiple peaks in the specific heat. The melting temperature for the longer chain is lower than the shorter chains. Multiple peaks are absent in the longer chain. The opening profile of the DNA of different lengths reveals the difference in the breaking of the bonds and opening pathway of the different systems.

The present study is an initial attempt to understand the complex dynamics of the crowders on the stability of the DNA molecule. There will be many possibilities of distributing the molecular crowders over the molecule. The computation of the melting temperature of DNA for all the distributions is a mammoth task. We have presented the study for the few distributions, which helps us understand the behaviour of DNA molecules in the presence of molecular crowders. How the time scale of the presence of a crowder at a particular location along the chain affect the melting of the DNA molecule will be part of our future studies.

\acknowledgements

We acknowledge the financial support from the Department of Science and Technology, New Delhi (EMR/2017/002451).

\bibliography{ref_crowd.bib}

\end{document}